\documentclass[a4paper,12pt, epsfig]{article}
\pdfoutput=1
\usepackage{epsfig}
\usepackage{epstopdf}
\usepackage{graphicx}
\usepackage{ifthen}
\usepackage{feynmp}
\usepackage{amsmath}
\usepackage[psamsfonts]{amssymb}
\usepackage{euscript}

\usepackage{latexsym}
\usepackage[arrow,matrix,curve]{xy}

\newskip\humongous \humongous=0pt plus 1000pt minus 1000pt

\newif\ifdtup

\def\,{\hspace{-.1cm}}
\def\hsp{,\hspace{.7cm}}

\def\fc#1#2 {\frac{n}{q}#1\frac{n}{q}#2}

\def\vpp{V^{\prime\prime}}
\def\vppp{V^{\prime\prime\prime}}

\renewcommand{\cos}{\textrm{cos}}
\renewcommand{\sin}{\textrm{sin}}

\renewcommand{\sinh}{\textrm{sinh}}
\renewcommand{\cosh}{\textrm{cosh}}
\renewcommand{\tanh}{\textrm{tanh}}
\newcommand{\sech}{\textrm{sech}}
\newcommand{\csch}{\textrm{csch}}

\def\exp#1{\hbox{\rm exp}\left(#1\right)}

\renewcommand{\theequation}{\arabic{section}.\arabic{equation}}
\renewcommand{\(}{\begin{equation}}
\renewcommand{\)}{end{equation} \vspace{-.05in}\linebreak}

\newcounter{saveeqn}
\newcounter{savealpheqn}

\newcommand{\alpheqn}{\setcounter{saveeqn}{\value{equation}}%
  \stepcounter{saveeqn}\setcounter{equation}{0}%
  \renewcommand{\theequation}{\mbox{\arabic{section}.\arabic{saveeqn}
\alph{equation}}}
  \renewcommand{\)}{\end{equation}}}
\def\part#1{\frac{\partial}{\partial{#1}}}%
\def\group#1{\refstepcounter{equation}\setcounter{saveeqn}
 {\value{equation}}%
  \label{#1}\setcounter{equation}{0}%
\renewcommand{\theequation}{\mbox{\arabic{section}.\arabic{saveeqn}
\alph{equation}}}
  \renewcommand{\)}{\end{equation}}}
\newcommand{\reseteqn}{\setcounter{equation}{\value{saveeqn}}%
  \renewcommand{\theequation}{\arabic{section}.\arabic{equation}}%
  \renewcommand{\)}{\end{equation}}}

\newcommand{\aalpheqn}{\setcounter{saveeqn}{\value{equation}}%
  \stepcounter{saveeqn}\setcounter{equation}{0}%
  \renewcommand{\theequation}{\mbox{
        \Alph{subsection}.\arabic{saveeqn}\alph{equation}}}
   \renewcommand{\)}{\end{equation}}}
\newcommand{\areseteqn}{\setcounter{equation}{\value{saveeqn}}%
  \renewcommand{\theequation}{\Alph{subsection}.\arabic{equation}}%
  \renewcommand{\)}{\end{equation}}}

\renewcommand{\thefootnote}{\alph{footnote}}
\renewcommand{\(}{\begin{equation}}
\renewcommand{\)}{\end{equation}}
\newcommand{\ba}{\begin{eqnarray}}
\newcommand{\ea}{\end{eqnarray}}
\newcommand{\cbp}{\mathop{\vtop{\ialign{##\crcr
   $\hfil\displaystyle{}\hfil$\crcr\noalign{\kern-13pt\nointerlineskip}
   \BIG{)}\hskip0pt\crcr\noalign{\kern3pt}}}}}
\newcommand{\pa}{\mathop{\vtop{\ialign{##\crcr

$\hfil\displaystyle{\oplus}\hfil$\crcr\noalign{\kern+1pt\nointerlineskip
}
   \hspace{.08in}$^{\alpha=0}$\hskip6pt\crcr\noalign{\kern3pt}}}}}
\renewcommand{\hsp}{,\hspace{.3in}}
\newcommand{\p}{^\prime}

\newcommand{\Z}{\ensuremath{\mathbb Z}}

\catcode`\@=11
\def\vereq#1#2{\lower3pt\vbox{\baselineskip1.5pt \lineskip1.5pt
\ialign{$\m@th#1\hfill##\hfil$\crcr#2\crcr\sim\crcr}}}
\catcode`\@=12

\renewcommand{\(}{\begin{equation}}
\renewcommand{\)}{\end{equation}}

\def\pin#1{\int \frac{d#1}{2\pi}}
\def\pink#1{\int \frac{d^{#1}k}{(2\pi)^{#1}}}
\def\sq#1#2{\sqrt{\frac{\omega_{#1}}{\omega_{#2}}}}
\def\bd#1{b^\dag_{k_{#1}}}
\def\bm#1{b_{-k_{#1}}}

\def\df{\mathcal{D}_f}

\newcommand{\beas}{\begin{eqnarray*}}
\newcommand{\eeas}{\end{eqnarray*}}

\newcommand{\bquo}{\begin{quote}}
\newcommand{\enqu}{\end{quote}}

\renewcommand{\Z}{{\mathbb Z}}

\def\ch{{\mathcal{H}}}
\def\co{{\mathcal{O}}}

\newcommand{\beq}{\begin{equation}}
\newcommand{\eeq}{\end{equation}}
\newcommand{\bea}{\begin{eqnarray}}
\newcommand{\eea}{\end{eqnarray}}

\newskip\humongous \humongous=0pt plus 1000pt minus 1000pt

\newif\ifdtup

\catcode`\@=11

\@addtoreset{equation}{section}

\def\@normalsize{\@setsize\normalsize{15pt}\xiipt\@xiipt
\abovedisplayskip 14pt plus3pt minus3pt%
\belowdisplayskip \abovedisplayskip
\abovedisplayshortskip \z@ plus3pt%
\belowdisplayshortskip 7pt plus3.5pt minus0pt}

\def\small{\@setsize\small{13.6pt}\xipt\@xipt
\abovedisplayskip 13pt plus3pt minus3pt%
\belowdisplayskip \abovedisplayskip
\abovedisplayshortskip \z@ plus3pt%
\belowdisplayshortskip 7pt plus3.5pt minus0pt
\def\@listi{\parsep 4.5pt plus 2pt minus 1pt
      \itemsep \parsep
      \topsep 9pt plus 3pt minus 3pt}}

\relax

\catcode`@=12

\catcode`\@=11

\def\section{\@startsection{section}{1}{\z@}{3.5ex plus 1ex minus  .2ex}{2.3ex plus .2ex}{\large\bf}}

\def\thesection{\arabic{section}}
\def\thesubsection{\arabic{section}.\arabic{subsection}}

\def\appendix{\setcounter{section}{0}
 \def\thesection{Appendix \Alph{section}}
 \def\thesubsection{\Alph{section}.\arabic{subsection}}
 \def\theequation{\Alph{section}.\arabic{equation}}}
\renewcommand{\theequation}{\arabic{section}.\arabic{equation}}

\begin{document}
\def\thefootnote{\fnsymbol{footnote}}
\def\thetitle{Normal Ordering Normal Modes}
\def\autone{Jarah Evslin}
\def\affa{Institute of Modern Physics, NanChangLu 509, Lanzhou 730000, China}
\def\affb{University of the Chinese Academy of Sciences, YuQuanLu 19A, Beijing 100049, China}

\begin{center}
{\large {\bf \thetitle}}

\bigskip

\bigskip

{\large \noindent  \autone{${}^{1,2}$} \footnote{jarah@impcas.ac.cn}}

\vskip.7cm

1) \affa\\
2) \affb\\

\end{center}

\begin{abstract}
\noindent
In a soliton sector of a quantum field theory, it is often convenient to expand the quantum fields in terms of normal modes.  Normal mode creation and annihilation operators can be normal ordered, and their normal ordered products have vanishing expectation values in the one-loop soliton ground state.  The Hamiltonian of the theory, however, is usually normal ordered in the basis of operators which create plane waves.  In this paper we find the Wick map between the two normal orderings.  For concreteness, we restrict our attention to Schrodinger picture scalar fields in 1+1 dimensions, although we expect that our results readily generalize beyond this case.  We find that plane wave ordered $n$-point functions of fields are sums of terms which factorize into $j$-point functions of zero modes, breather and continuum normal modes.  We find a recursion formula in $j$ and, for products of fields at the same point, we solve the recursion formula at all $j$. 


\end{abstract}

%
\setcounter{footnote}{0}
\renewcommand{\thefootnote}{\arabic{footnote}}

\section{Introduction}

In perturbation theory about a translation-invariant vacuum, it is customary to decompose the quantum fields into operators $a_p^\dag$ and $a_p$ which create and annihilate plane wave excitations.  The free vacuum is annihilated by $a_p$ and is the initial state in the perturbative expansion.  This perturbation theory is simplest when the Hamiltonian is normal ordered, so that all $a^\dag_p$ appear to the left of all $a_p$.

At the same leading order\footnote{Even at weak coupling quantum corrections can affect the existence itself of the solution \cite{delfino,davies}.}, the ground state of a quantum soliton is given by a coherent state formed by shifting the fields by the functions corresponding to their classical solutions \cite{hepp,taylor78,sato}.  The normal modes of the quantum soliton are, at linear order, described by quantum harmonic oscillators.  The one-loop ground state of the soliton sector consists of the tensor product of the ground states of these oscillators \cite{cahill76}.  If the fields are decomposed into the normal modes of the soliton, with operators $b^\dag_k$ and $b_k$ corresponding to the raising and lowering operators in the corresponding quantum harmonic oscillators, then the one-loop soliton ground state is, after the shift operator noted above, the state annihilated by all of the $b_k$.   

Again a sensible perturbation theory exists which describes the spectrum of the one soliton sector.  It is similar to that of the vacuum sector, except that treating zero modes requires special care \cite{christleecc,sakitacc,meduestato}.  In particular the calculation is simplest if the Hamiltonian is normal ordered by placing all $b^\dag_k$ to the left of $b_k$.  

However, the Hamiltonian is usually given normal ordered in terms of plane waves, as is convenient for vacuum sector perturbation theory.  Therefore the first step in soliton perturbation theory is to convert plane wave normal ordering to normal mode normal ordering.   The goal of the present note is to describe how this can be done in the case of a scalar field theory in 1+1 dimensions.  The fact that the theory is scalar and only in 1+1 dimensions does not appear to play a central role in our analysis, and so we expect that the approach in this paper can be trivially generalized to more complicated theories in more dimensions.

We find that the problem of converting plane wave normal ordering into normal mode normal ordering can be achieved in two steps.  First, as will be described in Sec.~\ref{decsez}, we show that plane wave normal ordered products of the form $:\phi^n(x):_a$ can be decomposed into sums of products of factors of the form $:\phi_M^j(x):_a$ where
\beq
\phi(x)=\sum_M \phi_M(x)
\eeq
is a decomposition into different kinds of normal modes, such as even and odd breather modes.  Next, in Sec.~\ref{mainsez}, we find that these factors can each be converted according to the Wick formula
\beq
:\phi^j_M(x):_a=\sum_{m=0}^{\lfloor{\frac{j}{2}}\rfloor}\frac{j!}{m!(j-2m)!}I_M^k(x):\phi^{j-2m}_M(x):_b
\eeq
where the contraction, except for the case of zero modes, is schematically
\beq
I_M(x)=\left\langle\frac{1}{2\omega_{k}}-\frac{1}{2\omega_p}\right\rangle
\eeq
with $\omega_k$ and $\omega_p$ the energy of a normal mode and plane wave respectively.  We begin in Sec.~\ref{initsez} with a review of our formalism.

\section{The Setup} \label{initsez}

\begin{table}
\begin{tabular}{|l|l|}
\hline
Operator&Description\\
\hline
$\phi(x),\ \pi(x)$&The real scalar field and its conjugate momentum\\
$a^\dag_p,\ a_p$&Creation and annihilation operators in plane wave basis\\
$b^\dag_k,\ b_k$&Creation and annihilation operators in normal mode basis\\
$b^\dag_{BE/BO},\ b_{BE/BO}$&Creation/annihilation operators for even/odd breather modes\\
$\phi_0,\ \pi_0$&Zero mode of $\phi(x)$ and $\pi(x)$ in normal mode basis\\
$::_a,\ ::_b$&Normal ordering with respect to $a$ or $b$ operators respectively\\
$S[]$&Symmetrization with respect to momenta\\
\hline
Indices&Description\\
\hline
$m$&Contractions\\
$i$&Breather modes\\
$I$&Bound states including both breather modes and the zero mode\\
$M$&Normal mode type: zero mode, breather or continuum mode\\
\hline
Hamiltonian&Description\\
\hline
$H$&The original Hamiltonian\\
$H^\prime$&$H$ with $\phi(x)$ shifted by soliton solution $f(x)$\\
$H_n$&The $\phi^n$ term in $H^\prime$\\
\hline
Symbol&Description\\
\hline
$f(x)$&The classical soliton solution\\
$\df$&Operator that translates $\phi(x)$ by the classical soliton solution\\
$g_B(x)$&The soliton linearized translation mode\\
$g_{BE,i}(x),\ g_{BO,i}(x)$&The $i$th even/odd breather mode\\
$g_k(x)$&Continuum normal mode\\
$p$&Momentum\\
$k_i$&The analog of momentum for soliton perturbations\\
$\omega_k,\ \omega_p$&The frequency corresponding to $k$ or $p$\\
$\tilde{g}$&Inverse Fourier transform of $g$\\
$I_M(x)$&Contraction arising from type $M$ normal mode\\
$N_k,\ N^M_k$&Plane wave normal ordered product of $k$ $a^\dag+a$ or $a^\dag_M+a_M$ factors\\
$B^M_n$&Normal mode normal ordered product of $n$\ $b^\dag\pm b$ factors\\
$\alpha_{nm},\ a_{nm}$&Dimensionful/less coefficients for $n$ field products with $m$ contractions\\
\hline
State&Description\\
\hline
$|K\rangle, \ |\Omega\rangle$&Kink and vacuum sector ground states\\
$\co|\Omega\rangle$&Translation of $|K\rangle$ by $\df^{-1}$\\
$\co_1|\Omega\rangle$&Translation of $|K\rangle$ at one loop by $\df^{-1}$\\
\hline

\end{tabular}
\caption{Summary of Notation}\label{notab}
\end{table}

In this section we will review the one loop description of kinks developed in Refs.~\cite{dhn2,rajaraman,aguirre} using the formalism developed in Refs.~\cite{cahill76,mekink,memassa}, which has the advantage that it resolves the ambiguity noted in Ref.~\cite{rebhan}.  The key elements of our notation are summarized in Table~\ref{notab}.

For concreteness, we consider a theory of a real scalar field $\phi(x)$ and its canonical momentum $\pi(x)$ in 1+1 dimensions, described by a Hamiltonian
\beq
H=\int dx \ch(x) \hsp
\ch(x)=\frac{1}{2}:\pi(x)\pi(x):_a+\frac{1}{2}:\partial_x\phi(x)\partial_x\phi(x):_a+\frac{M^2}{g^2}:V[g\phi(x)]:_a\label{hsq}
\eeq
where $M$ has dimensions of mass and $g$ has dimensions of action${}^{-1/2}$.  The perturbative expansion will be an expansion in $g^2\hbar$ and we will set $\hbar=1$.   The plane-wave normal-ordering $::_a$ will be defined momentarily.
 
We assume that the potential $V$ has degenerate minima so that the classical equations of motion admit a time-independent kink solution
\beq
\phi(x,t)=f(x).
\eeq
In the Schrodinger picture of the quantum theory, the translation operator
\beq
\df={\rm{exp}}\left(-i\int dx f(x)\pi(x)\right) \label{df}
\eeq
satisfies the identity \cite{mekink}
\beq
:F\left[\pi(x),\phi(x)\right]:_a\df=\df:F\left[\pi(x),\phi(x)+f(x)\right]:_a \label{fident}
\eeq
for any functional $F$ and maps the vacuum sector to the kink sector.   For example, the kink ground state may be written
\beq
|K\rangle=\df \co|\Omega\rangle
\eeq
where $|\Omega\rangle$ is the free scalar vacuum state and $\co$ may be calculated in perturbation theory.  As  $|K\rangle$ is a Hamiltonian eigenstate, $\co|\Omega\rangle$ is an eigenstate of its similarity transform
\bea
H\p&=&\df^{-1} H\df=Q_0+H_2+H_I\\
H_2&=&\frac{1}{2}\int dx\left[:\pi^2(x):_a+:\left(\partial_x\phi(x)\right)^2:_a+M^2V^{\prime\prime}[gf(x)]:\phi^2(x):_a\right]\nonumber
\eea
where $Q_0$ is the classical kink mass and $H_I$ consists of higher order terms in the $g$ expansion.   Note that $gf(x)$ is dimensionless and so contains no powers of $\hbar$ and so no powers of $g$.   

As $Q_0$ is $O(g^{-2})$ and $H_2$ is $O(g^0)$, these are the only terms which appear at one loop.  In particular the one loop kink ground state $\co_1|\Omega\rangle$ is an eigenstate of $H_2$.  To find it, one expands the fields in terms of the fixed frequency $\omega$ solutions $g(x)$ of the classical equations of motion for $H_2$
 \beq
 \phi(x,t)=e^{-i\omega t}g(x)\hsp
M^2 V^{\prime\prime}[f(x)]g(x)=\omega^2g(x)+g^{\prime\prime}(x). 
 \eeq
This is a wave equation for a particle in a potential and its solutions are the normal modes of the field theory in the kink background.  It generally has bound state and continuum solutions.  We will refer to even and odd bound state solutions as $g_{BE,i}(x)$ and $g_{BO,i}(x)$ respectively, where the index $i$ runs over distinct solutions if there is more than one.  There will always be an even bound state solution corresponding to the translation symmetry, which we call
\beq
g_B(x)=\frac{1}{\sqrt{Q_0}} f^\prime(x). 
\eeq
As it corresponds to a symmetry, it is a zero mode $\omega_B=0$.  The other bound state solutions correspond to breather modes.  Let $n_e$ and $n_o$ be the number of even and odd breather modes.  We will name the continuum states $g_k(x)$ where $k$ is defined by $\omega_k^2=k^2+m^2$ and the sign of $k$ is fixed by demanding that asymptotically it becomes the corresponding plane wave.   All of these solutions are clearly mutually orthogonal and we normalize them such that
\beq
\int dx g_{k_1} (x) g^*_{k_2}(x)=2\pi \delta(k_1-k_2)\hsp
\int dx |g_{B}(x)|^2=\int dx |g_{BE}(x)|^2=\int dx |g_{BO}(x)|^2=1. 
\eeq
We also impose
\beq
g(-x)=g^*(x). \label{congc}
\eeq
Their inverse Fourier transforms
\beq
\tilde{g}(p)=\int dx g(x) e^{ipx}
\eeq
satisfy the completeness relations
\beq
\sum_I\tilde{g}_{I}(p)\tilde{g}_{I}(q)+\pin{k}\tilde{g}_k(p)\tilde{g}_{-k}(q)=2\pi\delta(p+q) \label{comp}
\eeq
where $I$ runs over all $n_e+n_o+1$ bound state field labels $\{B,\{BE,i\},\{BO,i\}\}$.

One may expand the fields in terms of plane waves
\bea
\phi(x)&=&\pin{p}\frac{1}{\sqrt{2\omega_p}}\left(a^\dag_p+a_{-p}\right) e^{-ipx}\hsp
\omega_p=\sqrt{m^2+p^2}\label{pwexp}\\
 \pi(x)&=&i\pin{p}\sqrt{\frac{\omega_p}{2}}\left(a^\dag_p-a_{-p}\right) e^{-ipx}
\nonumber
\eea
or in terms of normal modes
\bea
\phi(x)&=&\sum_I \phi_{I}(x) +\phi_{C}(x)\hsp
\pi(x)=\sum_I \pi_{I}(x)+\pi_{C}(x) \label{nmexp}\\
\phi_B(x)&=&\phi_0 g_B(x)\hsp\phi_{BE,i}(x)=\frac{1}{\sqrt{2\omega_{BE,i}}}\left(b_{BE,i}^\dag+b_{BE,i}\right) g_{BE,i}(x)\nonumber\\
\phi_{BO,i}(x)&=&\frac{1}{\sqrt{2\omega_{BO,i}}}\left(b_{BO,i}^\dag-b_{BO,i}\right) g_{BO,i}(x)\hsp
\phi_C(x)=\pin{k}\frac{1}{\sqrt{2\omega_k}}\left(b_k^\dag+b_{-k}\right) g_k(x)\nonumber\\
\pi_B(x)&=&\pi_0 g_B(x)\hsp\pi_{BE,i}(x)=i\sqrt{\frac{\omega_{BE,i}}{2}}\left(b_{BE,i}^\dag - b_{BE,i}\right) g_{BE,i}(x)
\nonumber\\
\pi_{BO,i}(x)&=&i\sqrt{\frac{\omega_{BO,i}}{2}}\left(b_{BO,i}^\dag + b_{BO,i}\right) g_{BO,i}(x)\hsp
\pi_C(x)=i\pin{k}\sqrt{\frac{\omega_k}{2}}\left(b_k^\dag - b_{-k}\right) g_k(x).
\nonumber
\eea
Normal ordering may be defined with respect to either decomposition.  Plane wave normal ordering $::_a$ places all $a^\dag$ to the left of each $a$.  Normal mode normal ordering $::_b$ places all $b^\dag$ and $\phi_0$ on the left of all $b$ and $\pi_0$.

From the canonical algebra satisfied by $\phi(x)$ and $\pi(x)$ one easily finds the algebra satisfied by their components
\bea
[a_p,a_q^\dag]&=&2\pi\delta(p-q)\hsp
[\phi_0,\pi_0]=i\hsp
[b_{BE,i},b^\dag_{BE,j}]=\delta_{ij}\\
&&[b_{BO,i},b^\dag_{BO,j}]
=\delta_{ij}\hsp
[b_{k_1},b^\dag_{k_2}]=2\pi\delta(k_1-k_2)\nonumber
\eea
with all other commutators within each decomposition vanishing.  Finally one may simplify $H_2$
\beq
H_2=Q_1+\frac{\pi_0^2}{2}+\sum_{i=1}^{n_e} \omega_{BE,i}b^\dag_{BE,i}b_{BE,i}+\sum_{i=1}^{n_o} \omega_{BO,i}b^\dag_{BO,i}b_{BO,i}+\pin{k}\omega_k b^\dag_k b_k
\eeq
where $Q_1$ is the one-loop correction to the kink energy.  One recognizes this system as the sum of a free quantum mechanical particle with position $\phi_0$ and momentum $\pi_0$ plus an infinite set of quantum harmonic oscillators.  The one-loop vacuum therefore is annihilated by $\pi_0$ and also by all operators $b$
\beq
\pi_0\co_1|\Omega\rangle=b_{BE,i}\co_1|\Omega\rangle=b_{BO,i}\co_1|\Omega\rangle=b_k\co_1|\Omega\rangle=0.
\eeq
This means that normal mode normal ordered operators $:A:_b$, with vanishing c-number component, have vanishing expectation values at one loop
\beq
\langle\Omega|\co_1^\dag:A:_b\co_1|\Omega\rangle=0.
\eeq
This is one motivation for considering normal mode normal ordering.  Another is that it allows an efficient computation of states and energies beyond one loop \cite{meduestato}.  

\section{Factorization} \label{decsez}

\subsection{Factorization}

The Hamiltonian $H$ is plane wave normal ordered and a similarity transform by $\df$ preserves the normal ordering \cite{mekink}.  Therefore the Hamiltonian $H\p$ is also plane wave normal ordered.  However for several applications, normal mode normal ordering is most efficient.  In this paper we will study how to relate the two.  

The Hamiltonian $H\p$, at $n$th order, for $n>2$ is
\beq
H_n=\frac{M^2g^{n-2}}{n!}V^{(n)}[gf(x)]:\phi^n(x):_a
\eeq
where $V^{(n)}$ is the $n$th functional derivative of the potential $V$ with respect to its argument.  To calculate the soliton spectrum and energy corrections in perturbation theory, beginning with $\co_1|\Omega\rangle$, it is easiest to normal mode normal order the Hamiltonian.  The plane wave normal ordering is defined in terms of $a^\dag$ and $a$ and so, to evaluate these terms, we must use the plane wave expansion (\ref{pwexp})
\beq
:\phi^n(x):_a=\int \frac{d^np}{(2\pi)^n}\frac{\rm{exp}\left(-ix\sum_{i=1}^n p_i\right)}{\sqrt{2^n\omega_{p_1}\cdots\omega_{p_n}}}:\prod_{i=1}^n\left(a^\dag_{p_i}+a_{-p_i}\right):_a. \label{phin}
\eeq

To rewrite this in terms of normal mode operators, one need only insert (\ref{nmexp}) into the inverse of (\ref{pwexp}) to obtain the Bogoliubov transformations
\bea
a^\dag_p&=&\sum_I a^\dag_{I,p}+a^\dag_{C,p}\hsp
a_{-p}=\sum_I a_{I,-p}+a_{C,-p}\label{bog}\\
a^\dag_{B,p}&=&\tilde{g}_{B}(p)\left[ \sqrt{\frac{\omega_p}{2}}\phi_0-\frac{i}{\sqrt{2\omega_p}}\pi_0\right]\hsp
a_{B,-p}=\tilde{g}_{B}(p)\left[ \sqrt{\frac{\omega_p}{2}}\phi_0+\frac{i}{\sqrt{2\omega_p}}\pi_0\right].\nonumber\\
a^\dag_{BE,i,p}&=&\frac{\tilde{g}_{BE,i}(p)}{2}\left(\frac{\omega_p+\omega_{BE,i}}{\sqrt{\omega_p\omega_{BE,i}}}b_{BE,i}^\dag+\frac{\omega_p-\omega_{BE,i}}{\sqrt{\omega_p\omega_{BE,i}}}b_{BE,i}\right) \nonumber\\
a_{BE,i,-p}&=&\frac{\tilde{g}_{BE,i}(p)}{2}\left(\frac{\omega_p-\omega_{BE,i}}{\sqrt{\omega_p\omega_{BE,i}}}b_{BE,i}^\dag+\frac{\omega_p+\omega_{BE,i}}{\sqrt{\omega_p\omega_{BE,i}}}b_{BE,i}\right) \nonumber\\
a^\dag_{BO,i,p}&=&\frac{\tilde{g}_{BO,i}(p)}{2}\left(\frac{\omega_p+\omega_{BO,i}}{\sqrt{\omega_p\omega_{BO,i}}}b_{BO,i}^\dag+\frac{-\omega_p+\omega_{BO,i}}{\sqrt{\omega_p\omega_{BO,i}}}b_{BO,i}\right) \nonumber\\
a_{BO,i,-p}&=&\frac{\tilde{g}_{BO,i}(p)}{2}\left(\frac{\omega_p-\omega_{BO,i}}{\sqrt{\omega_p\omega_{BO,i}}}b_{BO,i}^\dag+\frac{-\omega_p-\omega_{BO,i}}{\sqrt{\omega_p\omega_{BO,i}}}b_{BO,i}\right) \nonumber\\
a^\dag_{C,p}&=&\pin{k}\frac{\tilde{g}_k(p)}{2}\left(\frac{\omega_p+\omega_k}{\sqrt{\omega_p\omega_k}}b_k^\dag+\frac{\omega_p-\omega_k}{\sqrt{\omega_p\omega_k}}b_{-k}\right) \nonumber\\
a_{C,-p}&=&\pin{k}\frac{\tilde{g}_k(p)}{2}\left(\frac{\omega_p-\omega_k}{\sqrt{\omega_p\omega_k}}b_k^\dag+\frac{\omega_p+\omega_k}{\sqrt{\omega_p\omega_k}}b_{-k}\right)\nonumber.
\eea
The key simplification comes from the fact that the modes from distinct oscillators commute with each other and they all commute with the zero modes.  Thus, after inserting (\ref{bog}) into (\ref{phin}), one can separate the modes of each oscillator and the zero modes
\beq
:\prod_{i=1}^n\left(a^\dag_{p_i}+a_{-p_i}\right):_a=\sum_{\{J^M|\cup_M J^M=[1,n]\}} \prod_M\left(:\prod_{i\in J^M}\left(a^\dag_{M,p_i}+a_{M,-p_i}\right):_a\right) \label{fattore}
\eeq
where $M$ runs over $\{B,\{BE,i\},\{BO,i\},C\}$, the $J^M$ are disjoint and their union is $[1,n]$.

For example, in the case of two point functions in the Sine-Gordon model, $n_e=n_0=0$ and $n=2$.  Thus $M$ runs over the labels $B$ and $C$ corresponding to the translation zero mode and the continuum.  $J^M$ runs over the four subsets of $\{1,2\}$, leading to four summands
\bea
:\prod_{i=1}^2\left(a^\dag_{p_i}+a_{-p_i}\right):_a&=&
:\prod_{i=1}^2\left(a^\dag_{B,p_i}+a_{B,-p_i}\right):_a+:\left(a^\dag_{B,p_1}+a_{B,-p_1}\right):_a:\left(a^\dag_{C,p_2}+a_{C,-p_2}\right):_a\nonumber\\
&&\hspace{-2cm}+:\left(a^\dag_{B,p_2}+a_{B,-p_2}\right):_a:\left(a^\dag_{C,p_1}+a_{C,-p_1}\right):_a+:\prod_{i=1}^2\left(a^\dag_{C,p_i}+a_{C,-p_i}\right):_a. \label{es}
\eea

Note that in a local Hamiltonian, normal-ordered products appear in the combination (\ref{phin}) where this product is integrated over a kernel which is symmetric with respect to permutations of the $p_i$.  Thus only the symmetric part of the product contributes to the Hamiltonian.  This depends on the subsets $J^M$ only via their cardinalities $j_M=|J^M|$ which sum to $n$
\beq
{\rm{S}}\left[:\prod_{i=1}^n\left(a^\dag_{p_i}+a_{-p_i}\right):_a\right]=n! \sum_{\{j_M|\sum_M j_M=n\}} {\rm{S}}\left[\prod_M\left(\frac{1}{j_M!}:\prod_{i=1+\sum_{N=1}^{M-1}j_N}^{\sum_{N=1}^{M}j_N}\left(a^\dag_{M,p_i}+a_{M,-p_i}\right):_a\right)\right] \label{fattoresim}
\eeq
where $S$ symmetrizes all values of $p_i$.  Where the letter $M$ appears in the limits of the sum, it is understood that we have numbered the $n_o+n_e+2$ values of $M$ from $1$ to $n_o+n_e+2$.  The ordering chosen does not matter.  

For example, (\ref{es}) becomes
\bea
S\left[:\prod_{i=1}^2\left(a^\dag_{p_i}+a_{-p_i}\right):_a\right]&=&
S\left[:\prod_{i=1}^2\left(a^\dag_{B,p_i}+a_{B,-p_i}\right):_a\right]\\
&&\hspace{-3cm}+2S\left[:\left(a^\dag_{B,p_1}+a_{B,-p_1}\right):_a:\left(a^\dag_{C,p_2}+a_{C,-p_2}\right):_a\right]+S\left[:\prod_{i=1}^2\left(a^\dag_{C,p_i}+a_{C,-p_i}\right):_a\right] \nonumber
\eea
where the three terms correspond to $\{j_B=2,\ j_C=0\}$, $\{j_B=1,\ j_C=1\}$ and $\{j_B=0,\ j_C=2\}$.  To avoid clutter, below the operator $S$ will not be written explicitly, but we will write in the text when we symmetrize. 

 If we decompose $\phi(x)$ similarly to the plane wave operators
\beq
\phi(x)=\sum_M \phi_M(x)\hsp
\phi_M(x)=\pin{p}\frac{1}{\sqrt{2\omega_p}}\left(a^\dag_{M,p}+a_{M,-p}\right) e^{-ipx}
\eeq
then we can use (\ref{fattoresim}) to decompose 
\beq
:\phi^n(x):_a=n! \sum_{\{j_M|\sum_M j_M=n\}} \prod_M\left(\frac{1}{j_M!}:\phi^{j_M}_M(x):_a\right). \label{fattorephi}
\eeq
Note that the symmetrization is automatic here because of the symmetric kernel of the $p$ integration in (\ref{phin}).

The normal ordering on the right hand side of (\ref{fattore}) is defined to be whatever one obtains from (\ref{phin}) when all of the different oscillators are separated.  This is well-defined.  But is it a normal ordering?

\subsection{The Problem}

To simplify this question, let us restrict our attention momentarily to the case $n_e=n_o=0$, as in the Sine-Gordon theory.  The generalization to other values is trivial.  Clearly, whatever $::_a$ on the $a_{M}$ means, it is linear since the factorization above can be performed separately for each summand.  So consider one summand in (\ref{es})
\beq
:a^\dag_{B,p_1}a^\dag_{B,p_2}:_a.
\eeq
The simplest guess would be that $::_a$ places the $a_B^\dag$ on the left, and so the answer could be $a^\dag_{B,p_1}a^\dag_{B,p_2}$ or $a^\dag_{B,p_2}a^\dag_{B,p_1}$.  The trouble is that these are not equal because
\bea
[a^\dag_{B,p_1},a^\dag_{B,p_2}]&=&\left[\tilde{g}_{B}(p_1)\left( \sqrt{\frac{\omega_{p_1}}{2}}\phi_0-\frac{i}{\sqrt{2\omega_{p_1}}}\pi_0\right),\tilde{g}_{B}(p_2)\left( \sqrt{\frac{\omega_{p_2}}{2}}\phi_0-\frac{i}{\sqrt{2\omega_{p_2}}}\pi_0\right)\right]\nonumber\\
&=&\frac{1}{2}\left(\sq{p_1}{p_2}-\sq{p_2}{p_1}\right)\tilde{g}_{B}(p_1)\tilde{g}_{B}(p_2).
\eea
Similarly, in the case of continuum modes
\bea
[a^\dag_{C,p_1},a^\dag_{C,p_2}]&=&\pink{2}\left[\frac{\tilde{g}_{k_1}(p_1)}{2}\left(\frac{\omega_{p_1}+\omega_{k_1}}{\sqrt{\omega_{p_1}\omega_{k_1}}}b_{k_1}^\dag+\frac{\omega_{p_1}-\omega_{k_1}}{\sqrt{\omega_{p_1}\omega_{k_1}}}b_{-k_1}\right),\right.\\
&&\left.\frac{\tilde{g}_{k_2}(p_2)}{2}\left(\frac{\omega_{p_2}+\omega_{k_2}}{\sqrt{\omega_{p_2}\omega_{k_2}}}b_{k_2}^\dag+\frac{\omega_{p_2}-\omega_{k_2}}{\sqrt{\omega_{p_2}\omega_{k_2}}}b_{-k_2}\right)\right]\nonumber\\
&=&\frac{1}{2}\left(\sq{p_1}{p_2}-\sq{p_2}{p_1}\right) \pink{1}\tilde{g}_{k_1}(p_1)\tilde{g}_{-k_1}(p_2).
\nonumber
\eea
Thus the action of $::_a$ on $a^\dag_{M,p}$ and $a_{M,-p}$ is more complicated than simply putting all $a_M^\dag$ on the left, since their order matters.  This was not the case with the undecomposed plane wave oscillator modes because
\bea
[a^\dag_{p_1},a^\dag_{p_2}]&=&[a^\dag_{B,p_1},a^\dag_{B,p_2}]+[a^\dag_{C,p_1},a^\dag_{C,p_2}]\\
&=&\frac{1}{2}\left(\sq{p_1}{p_2}-\sq{p_2}{p_1}\right)\left(\tilde{g}_{B}(p_1)\tilde{g}_{B}(p_2)+ \pink{1}\tilde{g}_{k_1}(p_1)\tilde{g}_{-k_1}(p_2)\right)
\nonumber\\
&=&\frac{1}{2}\left(\sq{p_1}{p_2}-\sq{p_2}{p_1}\right)2\pi\delta(p_1+p_2)=0\nonumber
\eea
where we used the completeness relations (\ref{comp}) and the product of zero and a delta function vanishes at $p_1=p_2$ because this is the commutator of an operator with itself.

\noindent
{\bf{Conclusion:}} One may freely interchange the undecomposed plane wave mode operators $a^\dag$ and also $a$ inside of $::_a$, for example in Eq.~(\ref{phin}).  However, this shuffling fixes the order of the components $a^\dag_M$ and also $a_M$ in (\ref{fattore}).  In particular, the ordering of the components must be the same for all $M$, as this ordering is that chosen for the undecomposed operators.

Note that the symmetrized commutators vanish, and so this problem does not arise in the symmetrized products relevant to the computations of products of fields at the same point, as appear for example in the Hamiltonian.

\subsection{A Practical Convention}

In the previous subsection we learned that we need to make a choice.  We need to choose the ordering of the $a^\dag_{p_i}$ and also of the $a_{-p_i}$ in (\ref{phin}).  This choice does not affect our answer but it fixes the orderings of each component in (\ref{fattore}).  In this subsection we will choose an ordering which will facilitate the computations in the next section.

Let us define the shorthand
\beq
N_k(p_1\cdots p_k)=:\prod_{i=1}^k\left(a^\dag_{p_i}+a_{-p_i}\right):_a.
\eeq
We choose the ordering defined by
\beq
N_0=1\hsp
N_{k+1}(p_1\cdots p_{k+1})=a^\dag_{p_{k+1}}N_k(p_1\cdots p_k)+N_k(p_1\cdots p_k) a_{-p_{k+1}}.
\eeq
We remind the reader that the value of $N_k$ does not depend on this choice of ordering, as all $a^\dag$ commute with each other as do all $a$.  However it does affect the definition of the normal ordering of the components.

Our strategy will be the following.  First we will guess a formula for the normal ordering of the components
\beq
N^M_k(p_1\cdots p_k)=:\prod_{i=1}^k\left(a^\dag_{M,p_i}+a_{M,-p_i}\right):_a. \label{nm}
\eeq
Then we will show that, using the factorization formula (\ref{fattore}) the guess yields the correct value of $N_k$.  Recall that our definition of $::_a$ on components is that it satisfies (\ref{fattore}) and so once we have shown this, we will have verified that our guess indeed satisfies the definition and so corresponds to a valid convention.

Our guess is
\beq
N^M_0=1\hsp
N^M_{k+1}(p_1\cdots p_{k+1})=a^\dag_{M,p_{k+1}}N^M_k(p_1\cdots p_k)+N^M_k(p_1\cdots p_k) a_{M,-p_{k+1}}. \label{guess}
\eeq
The factorization formula (\ref{fattore}) in the case $n_e=n_o=0$ is
\beq
N_k(p_1\cdots p_k)=\sum_{J\subset[1,k]} N_{|J|}^B(p_J)N_{k-|J|}^C(p_{[1,k]\setminus J}). \label{fac2}
\eeq
Here we have adopted the shorthand $p_S$ for the ordered set of all $p_j$ with $j\in S$.  The ordering is just the ascending order, since that appeared on the left hand side of the equation.   We need to show that our guess (\ref{guess}) satisfies (\ref{fac2}).

Our proof will be by induction.  The base case, $k=0$ is trivial as the only term in the sum is $J=\varnothing$ and so (\ref{fac2}) becomes $1=1$.  Next assume that (\ref{fac2}) is satisfied for some value of $k$ and define
\beq
\hat{N}_{k+1}(p_1\cdots p_{k+1})=\sum_{J\subset[1,k+1]} N_{|J|}^B(p_J)N_{k+1-|J|}^C(p_{[1,k+1]\setminus J})
\eeq
where the right hand side is defined using (\ref{guess}).  We need to prove that $\hat{N}=N$ to complete the induction.

Each $J$ either does or does not contain the element $\{k+1\}$ and so we may respectively divide the sum in two parts, redefining the dummy set $J$ in the first sum by removing $\{k+1\}$
\bea
\hat{N}_{k+1}(p_1\cdots p_{k+1})&=&\sum_{J\subset[1,k]} N_{|J|+1}^B(p_J,p_{k+1})N_{k-|J|}^C(p_{[1,k]\setminus J})\\
&&+\sum_{J\subset[1,k]} N_{|J|}^B(p_J)N_{k+1-|J|}^C(p_{[1,k]\setminus J},p_{k+1})\nonumber\\
&=&\sum_{J\subset[1,k]} \left(a^\dag_{B,p_{k+1}}N_{|J|}^B(p_J,p_{k})+N_{|J|}^B(p_J,p_{k})a_{B,-p_{k+1}}\right) N_{k-|J|}^C(p_{[1,k]\setminus J})
\nonumber\\
&+&\sum_{J\subset[1,k]} N_{|J|}^B(p_J,p_{k})\left(a^\dag_{C,p_{k+1}}N_{k-|J|}^C(p_{[1,k]\setminus J})+N_{k-|J|}^C(p_{[1,k]\setminus J})a_{C,-p_{k+1}}\right)
\nonumber\\
&=&a^\dag_{p_{k+1}}N_k(p_1\cdots p_k)+N_k(p_1\cdots p_k)a_{-p_{k+1}}=N_{k+1}(p_1\cdots p_{k+1})\nonumber
\eea
completing the induction.

In summary, we have shown that if we adopt the definition (\ref{guess}) for the plane wave normal ordering of component fields $a^\dag_M$ and $a_M$, then the factorization formula (\ref{fac2}) is satisfied and so these components $N^M_k$ can be assembled to determine the plane wave normal ordered product $N_k$ of the undecomposed operators.  Although our proof was for the case with no breather modes $n_e=n_o=0$, the equation (\ref{guess}) works in general and indeed the proof can be trivially generalized to show the compatibility of (\ref{guess}) and (\ref{fattore}).

\section{Recursion Formulas} \label{mainsez}

\subsection{Zero Modes}
Define the coefficients $\alpha_{nm}$ by
\beq
N^B_n(p_1\cdots p_n)=\left(\prod_{i=1}^n\sqrt{2\omega_{p_i}}\tilde{g}_B(p_i)\right)\sum_{m=0}^{\lfloor{\frac{n}{2}}\rfloor}\alpha_{nm}\phi_0^{n-2m}
\eeq
where $N^B_n$ was defined  in (\ref{nm}).  Then using (\ref{guess}) we can find the next product
\bea
N^B_{n+1}(p_1\cdots p_{n+1})&=&\left(\prod_{i=1}^n\sqrt{2\omega_{p_i}}\tilde{g}_B(p_i)\right)\sum_{m=0}^{\lfloor{\frac{n}{2}}\rfloor}\alpha_{nm}\left(a^\dag_{B,p_{n+1}}\phi_0^{n-2m}+\phi_0^{n-2m}a_{B,-p_{n+1}}\right)\\
&=&\frac{1}{2}\left(\prod_{i=1}^{n+1}\sqrt{2\omega_{p_i}}\tilde{g}_B(p_i)\right)\sum_{m=0}^{\lfloor{\frac{n}{2}}\rfloor}\alpha_{nm}\left[\left(\phi_0-\frac{i}{\omega_{p_{n+1}}}\pi_0\right)\phi_0^{n-2m}\right.\nonumber\\
&&\left.+\phi_0^{n-2m}\left(\phi_0+\frac{i}{\omega_{p_{n+1}}}\pi_0\right)\right]
\nonumber\\
&=&\left(\prod_{i=1}^{n+1}\sqrt{2\omega_{p_i}}\tilde{g}_B(p_i)\right)\sum_{m=0}^{\lfloor{\frac{n}{2}}\rfloor}\alpha_{nm}\left(\phi_0^{n-2m+1}-\frac{1}{2\omega_{p_{n_1}}}\phi_0^{n-2m-1}\right).
\eea
Dividing through by the product on the left one finds
\beq
\sum_{m=0}^{\lfloor{\frac{n+1}{2}}\rfloor}\alpha_{n+1,m}\phi_0^{n-2m+1}=\sum_{m=0}^{\lfloor{\frac{n}{2}}\rfloor}\alpha_{nm}\left(\phi_0^{n-2m+1}-\frac{n-2m}{2\omega_{p_{n+1}}}\phi_0^{n-2m-1}\right).
\eeq
Finally matching terms with the same power of $\phi_0$ we arrive at the recursion relation
\beq
\alpha_{n+1,m}=\alpha_{nm}-\frac{n-2m+2}{2\omega_{p_{n+1}}}\alpha_{n,m-1} \label{brec}
\eeq
which, together with the initial condition $\alpha_{0m}=\delta_{m,0}$ fixes all of the coefficients $\alpha$.

The recursion relation (\ref{brec}) has a simple interpretation in terms of a Wick's theorem.  $m$ is the number of contractions.  The $(n+1)$st operator may either not contract, leading to the first term on the right hand side, or else it may contract.  If it does contract, since there are $m$ contractions in all, the first $n$ operators have $m-1$ contractions.  Therefore the $n+1$st operator may contract with any one of the $n-2(m-1)$ uncontracted operators, yielding the factor of $n-2m+2$ in the second term.  Each contraction yields a factor of $-1/(2\omega_{p_{n+1}})$.  Note that this contraction factor is not symmetric with respect to a permutation of the $p_i$, since it depends only on the $p_i$ with the highest value of $i$ among the two contracted operators, which is $p_{n+1}$.  

\subsection{Solving the Recursion Formula}
Recall that to compute the Hamiltonian we only need the symmetrized $N_n$.  In this case the choice of $\omega_{p_i}$ is irrelevant, it is only important that no $N$ have two $\omega_{p_i}$ with the same $i$.  Said differently, adding an antisymmetric piece to $N$ will not change the symmetrized $N$ and so will not change $H$.  We can thus shift $N$ to be of the form
\beq
N^B_n(p_1\cdots p_n)=\left(\prod_{i=1}^n\sqrt{2\omega_{p_i}}\tilde{g}_B(p_i)\right)\sum_{m=0}^{\lfloor{\frac{n}{2}}\rfloor}a_{nm}\phi_0^{n-2m}\prod_{i=1}^m\left(\frac{-1}{2\omega_{p_i}}\right) \label{banz}
\eeq
where $a$ is a pure number which simply counts the number of ways to make $m$ contractions.  $a$ satisfies the recursion relation
\beq
a_{n+1,m}=a_{nm}+(n-2m+2)a_{n,m-1}. \label{arec}
\eeq
As the contractions are interchangeable, $a_{nm}$ contains a factor of $1/m!$.  This is multiplied by the number of choices for the $j$th contraction, which is ${n-2j+2}\choose{2}$, for each $j$ from $1$ to $m$.  In all one finds
\beq
a_{nm}=\frac{1}{m!}\prod_{j=1}^m{{n-2j+2}\choose{2}}=\frac{1}{2^m}\frac{n!}{m!(n-2m)!}. \label{aeq}
\eeq

In the case with no breathers, the decomposition of the fields (\ref{fattorephi}) becomes
\beq
:\phi^n(x):_a=\sum_{j=0}^n {n\choose j} :\phi^{j}_B(x):_a :\phi^{n-j}_C(x):_a. \label{fattorephi2}
\eeq
Assembling the results above, we have evaluated the first factor in (\ref{fattorephi2})
\bea
 :\phi^{j}_B(x):_a &=&\int\frac{d^jp}{(2\pi)^j}\frac{e^{-ix\sum_ip_i}}{\sqrt{2^j\omega_{p_1}\cdots\omega_{p_j}}}N^B_j(p_1\cdots p_j)\label{bmaster}\\
 &=&\sum_{m=0}^{\lfloor{\frac{j}{2}}\rfloor}\frac{1}{2^m}\frac{j!}{m!(j-2m)!}\phi_0^{j-2m}\int\frac{d^jp}{(2\pi)^j}e^{-ix\sum_ip_i}\left(\prod_{i=1}^j\tilde{g}_B(p_i)\right)\prod_{i=1}^m\left(\frac{-1}{2\omega_{p_i}}\right)
 \nonumber\\
 &=&\sum_{m=0}^{\lfloor{\frac{j}{2}}\rfloor}\frac{j!}{m!(j-2m)!}g_B^{j-2m}(x)I_B^m(x)\phi_0^{j-2m}=\sum_{m=0}^{\lfloor{\frac{j}{2}}\rfloor}\frac{j!}{m!(j-2m)!}I_B^m(x)\phi_B^{j-2m}(x)
 \nonumber
 \eea
where we have introduced the contraction factor
\beq
I_B(x)=\frac{1}{2}g_B(x)\hat{g}_B(x)\hsp \hat{g}_B(x)=-\pin{p}e^{-ipx}\frac{\tilde{g}_B(p)}{2\omega_p}.
\eeq

\subsection{Example: The Sine-Gordon Theory}

In the Sine-Gordon theory the interaction Hamiltonian density in $H^\prime$ is \cite{sg}
\beq
\ch_I=\frac{m^2}{\sqrt{\lambda}}\sin(\sqrt{\lambda}f(x)) \sum_{n=1}^{\infty}\frac{(-\lambda)^n}{(2n+1)!} :\phi^{2n+1}(x):_a-\frac{m^2}{\lambda}\cos(\sqrt{\lambda}f(x))\sum_{n=2}^{\infty}\frac{(-\lambda)^n}{2n!} :\phi^{2n}(x):_a.
\eeq
The contribution arising from bound states is
\bea
\ch_B&=&-\frac{m^2}{\lambda}\cos(\sqrt{\lambda}f(x))h_e+\frac{m^2}{\sqrt{\lambda}}\sin(\sqrt{\lambda}f(x))h_o\label{chb}\\
h_e&=&\sum_{n=2}^{\infty}\frac{(-\lambda)^n}{2n!} :\phi_B^{2n}(x):_a\hsp
h_o= \sum_{n=1}^{\infty}\frac{(-\lambda)^n}{(2n+1)!} :\phi_B^{2n+1}(x):_a.\nonumber
\eea
Using (\ref{bmaster}) the plane wave normal ordering may be evaluated explicitly
\beq
h_e=\sum_{n=2}^{\infty}(-\lambda)^n \sum_{m=0}^{n}\frac{1}{m!(2n-2m)!}I_B^m(x)\phi_B^{2n-2m}(x).
\eeq

To simplify this sum, we will include the terms at $n=0$ and $n=1$, which are present in the Hamiltonian although they are not the only terms at their orders.  These terms only affect the noninteracting part of the Hamiltonian, which is known to be the Poschl-Teller Hamiltonian.  So we redefine
\bea
h_e&=&\sum_{n=0}^{\infty}(-\lambda)^n \sum_{m=0}^{n}\frac{1}{m!(2n-2m)!}I_B^m(x)\phi_B^{2n-2m}(x)\\
&=&\sum_{p=0}^{\infty} \sum_{m=0}^{\infty} \frac{(-\lambda)^{p+m}}{m!(2p)!}I_B^m(x)\phi_B^{2p}(x)=\cos\left(\sqrt{\lambda}\phi_B(x)\right)\exp{-\lambda I_B(x)}.
\nonumber
\eea
Similarly, including the $n=0$ term,
\bea
h_o&=& \sum_{n=0}^{\infty}(-\lambda)^n\sum_{m=0}^{n}\frac{1}{m!(2n-2m+1)!}I_B^m(x)\phi_B^{2n-2m+1}(x)\\
&=& \sum_{p=0}^{\infty}\sum_{m=0}^{\infty}\frac{(-\lambda)^{p+m}}{m!(2p+1)!}I_B^m(x)\phi_B^{2p+1}(x)
=\frac{1}{\sqrt{\lambda}}\sin\left(\sqrt{\lambda}\phi_B(x)\right)\exp{-\lambda I_B(x)}.\nonumber
\eea

Substituting this back into (\ref{chb}) we find
\beq
\ch_B=-\frac{m^2}{\lambda}\cos\left(\sqrt{\lambda}\left(\phi_B(x)+f(x)\right)\right)\exp{-\lambda I_B(x)}.
\eeq
This has a straightforward interpretation.  The combination $\phi_B(x)+f(x)$ is just the $\df$ translated field, brutally truncated to the zero mode part.  The prefactor and the cosine term are thus just the original Sine-Gordon action, translated and truncated.  However we see that the plane wave normal ordering is now gone, indeed it was our goal to eliminate it, and instead there is an exponential of a contraction term.  Thus plane wave normal ordering is equivalent to multiplication by the exponent of the bound state contraction.  Of course only the bound state contraction appeared because we have truncated our Hamiltonian by only considering the bound component of the field.  Our result is trivially normal mode normal ordered as it only involves the operator $\phi_0$.

More generally we may expect the exponential to include the sum of the contractions of the various normal modes
\beq
\ch_I=-\frac{m^2}{\lambda}:\cos\left(\sqrt{\lambda}\left(\phi(x)+f(x)\right)\right):_b\exp{-\lambda \sum_M I_M(x)}. \label{cong}
\eeq

\subsection{Odd Breathers}

Similarly to the plane wave ordered products $N_n(p)$ we will define the normal mode ordered products
\beq
B^{BO}_n=:\left(b^\dag_{BO}-b_{BO}\right)^n:_b.
\eeq
Our goal in this subsection is to learn how to expand $N_n(p)$ in terms of $B^{BO}_n(k)$.

Using the identity 
\beq
B^{BO}_n=\sum_{k=0}^n (-1)^k{n\choose k} b_{BO}^{\dag n-k} b_{BO}^k
\eeq
one readily derives the anticommutator
\beq
\{b^\dag_{BO}-b_{BO},B^{BO}_n\}=2B^{BO}_{n+1}-2nB^{BO}_{n-1}
\eeq
and the commutator
\beq
\left[b^\dag_{BO}+b_{BO},B^{BO}_n\right]=2nB^{BO}_{n-1}
\eeq
which will be useful momentarily.

Proceeding as for the zero mode, we define coefficients $\alpha_{nm}$ by 
\beq
N^{BO}_n(p_1\cdots p_n)=\left(\prod_{i=1}^n\sq{p_i}{BO}\tilde{g}_{BO}(p_i)\right)\sum_{m=0}^{\lfloor{\frac{n}{2}}\rfloor}\alpha_{nm}B^{BO}_{n-2m}.
\eeq
Then using (\ref{guess})
\bea
N^{BO}_{n+1}(p_1\cdots p_{n+1})&=&\left(\prod_{i=1}^n\sq{p_i}{BO}\tilde{g}_{BO}(p_i)\right)\sum_{m=0}^{\lfloor{\frac{n}{2}}\rfloor}\alpha_{nm}\left(a^\dag_{BO,p_{n+1}}B^{BO}_{n-2m}+B^{BO}_{n-2m}a_{BO,-p_{n+1}}\right)\nonumber\\
&=&\frac{1}{2}\left(\prod_{i=1}^{n+1}\sq{p_i}{BO}\tilde{g}_{BO}(p_i)\right)\nonumber\\
&&\times\sum_{m=0}^{\lfloor{\frac{n}{2}}\rfloor}\alpha_{nm}\left(\{b^\dag_{BO}-b_{BO},B^{BO}_{n-2m}\}+\frac{\omega_{BO}}{\omega_{p_{n+1}}}[b^\dag_{BO}+b_{BO},B^{BO}_{n-2m}]\right)\nonumber\\
&=&\left(\prod_{i=1}^{n+1}\sq{p_i}{BO}\tilde{g}_{BO}(p_i)\right)\nonumber\\
&&\times\sum_{m=0}^{\lfloor{\frac{n}{2}}\rfloor}\alpha_{nm}\left(B^{BO}_{n-2m+1}+(n-2m)\left(-1+\frac{\omega_{BO}}{\omega_{p_{n+1}}}\right)B^{BO}_{n-2m-1}\right)
\eea
and so
\beq
\sum_{m=0}^{\lfloor{\frac{n}{2}}\rfloor}\alpha_{nm}\left(B^{BO}_{n-2m+1}+(n-2m)\left(-1+\frac{\omega_{BO}}{\omega_{p_{n+1}}}\right)B^{BO}_{n-2m-1}\right)=\sum_{m=0}^{\lfloor{\frac{n+1}{2}}\rfloor}\alpha_{n+1,m}B^{BO}_{n-2m+1}.
\eeq
Matching coefficients we obtain the recursion relation
\beq
\alpha_{n+1,m}=\alpha_{nm}+(n-2m+2)\left(-1+\frac{\omega_{BO}}{\omega_{p_{n+1}}}\right)\alpha_{n,m-1}. \label{borec}
\eeq

So far we have not used symmetrization, and so our recursion relation may be applied to computing any $n$-point function.  Again, for calculating $n$-point functions at the same point, as in our interaction terms, we may shift $N^{BO}$ by an operator which vanishes when symmetrized
\beq
N^{BO}_n(p_1\cdots p_n)=\left(\prod_{i=1}^n\sq{p_i}{BO}\tilde{g}_{BO}(p_i)\right)\sum_{m=0}^{\lfloor{\frac{n}{2}}\rfloor}a_{nm}B^{BO}_{n-2m}\prod_{i=1}^m\left(-1+\frac{\omega_{BO}}{\omega_{p_i}}\right).
\eeq
Note that the product on the right can be rewritten
\beq
\prod_{i=1}^m\left(-1+\frac{\omega_{BO}}{\omega_{p_i}}\right)=\left(2\omega_{BO}\right)^m\prod_{i=1}^m\left(-\frac{1}{2\omega_{BO}}+\frac{1}{2\omega_{p_i}}\right)
\eeq
so that it resembles the contraction terms in (\ref{banz}).  Proceeding as above, the recursion formula satisfied by the $a_{nm}$ is again (\ref{arec}) and so the $a_{nm}$ are given by (\ref{aeq}).

\bea
 :\phi^{j}_{BO}(x):_a &=&\int\frac{d^jp}{(2\pi)^j}\frac{e^{-ix\sum_ip_i}}{\sqrt{2^j\omega_{p_1}\cdots\omega_{p_j}}}N^{BO}_j(p_1\cdots p_j)\\
 &=&\sum_{m=0}^{\lfloor{\frac{j}{2}}\rfloor}\frac{1}{2^m}\frac{j!}{m!(j-2m)!}B^{BO}_{j-2m}\int\frac{d^jp}{(2\pi)^j}e^{-ix\sum_ip_i}(2\omega_{BO})^{(2m-j)/2}\left(\prod_{i=1}^j\tilde{g}_{BO}(p_i)\right)\nonumber\\
 &&\times\prod_{i=1}^m\left(-\frac{1}{2\omega_{BO}}+\frac{1}{2\omega_{p_i}}\right)
 \nonumber\\
 &=&\sum_{m=0}^{\lfloor{\frac{j}{2}}\rfloor}\frac{j!}{m!(j-2m)!}I_{BO}^m(x)\frac{g_{BO}^{j-2m}(x)B^{BO}_{j-2m}}{(2\omega_{BO})^{(j-2m)/2}}\nonumber\\
 &=&\sum_{m=0}^{\lfloor{\frac{j}{2}}\rfloor}\frac{j!}{m!(j-2m)!}:\phi_{BO}^{j-2m}(x):_bI_{BO}^m(x)
 \nonumber
 \eea
where  we have defined the contraction factor
\beq
I_{BO}(x)=\frac{1}{2}g_{BO}(x)\hat{g}_{BO}(x)\hsp \hat{g}_{BO}(x)=\pin{p}e^{-ipx}\tilde{g}_{BO}(p)\left(-\frac{1}{2\omega_{BO}}+\frac{1}{2\omega_p}\right).
\eeq
The contraction factor is similar to $I_B(x)$ except that the contraction contains two terms $1/(2\omega_{BO})$ and $1/(2\omega_p)$ with a relative sign.  These are respectively the contraction arising from the normal mode normal ordering and the plane wave normal ordering.  In the case of $I_B(x)$ the normal mode normal ordering was fundamentally different, as it was a rule for the placement of the canonical variables $\phi_0$ and $\pi_0$ and not for the oscillator modes.

The occurrence of a difference of contractions in $I_{BO}$ is reminiscent of the general contraction defined in Ref.~\cite{diosi}.  The appearance of contractions in an exponential in (\ref{cong}) is also similar to the generalized Wick's theorem postulated there.  It would be useful to understand this connection more precisely, as the generalized Wick's theorem may provide a simple extension of our results to more complicated and interesting models.

\subsection{Even Breathers}

The normal ordering of even breathers is identical to that of odd breathers except for a few sign differences.  Defining
\beq
B^{BE}_n=:\left(b^\dag_{BE}+b_{BE}\right)^n:_b
\eeq
and using the identity
\beq
B^{BE}_n=\sum_{k=0}^n {n\choose k} b_{BE}^{\dag n-k} b_{BE}^k
\eeq
one finds
\beq
\{b^\dag_{BE}+b_{BE},B^{BE}_n\}=2B^{BE}_{n+1}+2nB^{BE}_{n-1}\hsp
\left[b^\dag_{BE}-b_{BE},B^{BE}_n\right]=-2nB^{BE}_{n-1}.
\eeq
Then defining
\beq
N^{BE}_n(p_1\cdots p_n)=\left(\prod_{i=1}^n\sq{p_i}{BE}\tilde{g}_{BO}(p_i)\right)\sum_{m=0}^{\lfloor{\frac{n}{2}}\rfloor}\alpha_{nm}B^{BE}_{n-2m}.
\eeq
The same computation as in the odd case yields the recursion relation
\beq
\alpha_{n+1,m}=\alpha_{nm}+(n-2m+2)\left(1-\frac{\omega_{BO}}{\omega_{p_{n+1}}}\right)\alpha_{n,m-1}. \label{berec}
\eeq
Comparing (\ref{borec}) and (\ref{berec}) one sees that the contractions of even and odd breathers differ by an overall sign.

In the symmetric case one may shift $N^{BE}$ to
\beq
N^{BE}_n(p_1\cdots p_n)=\left(\prod_{i=1}^n\sq{p_i}{BE}\tilde{g}_{BE}(p_i)\right)\sum_{m=0}^{\lfloor{\frac{n}{2}}\rfloor}a_{nm}B^{BE}_{n-2m}\prod_{i=1}^m\left(1-\frac{\omega_{BE}}{\omega_{p_i}}\right).
\eeq
where $a_{bm}$ again satisfies (\ref{arec}) and so we conclude that
\beq
 :\phi^{j}_{BE}(x):_a=\sum_{m=0}^{\lfloor{\frac{j}{2}}\rfloor}\frac{j!}{m!(j-2m)!}:\phi_{BE}^{j-2m}(x):_bI_{BE}^m(x)
 \eeq
where  we have defined the contraction factor
\beq
I_{BE}(x)=\frac{1}{2}g_{BE}(x)\hat{g}_{BE}(x)\hsp \hat{g}_{BE}(x)=\pin{p}e^{-ipx}\tilde{g}_{BE}(p)\left(\frac{1}{2\omega_{BE}}-\frac{1}{2\omega_p}\right).
\eeq
The relative sign in the recursion relation has indeed translated into a relative sign in the contraction factor with respect to $I_{BO}$.  As $g_{BO}(x)$ is imaginary and $g_{BE}(x)$ is real due to our convention (\ref{congc}), the relative sign may be absorbed by taking the complex conjugate of $g(x)$ in the definition of $I(x)$.  We will now see in the continuum case that this definition arises quite naturally.

\subsection{Continuum Modes}

Define
\beq
B^{C}_n(k_1\cdots k_n)=:\prod_{i=1}^n\left(\frac{b^\dag_{k_i}+b_{-k_i}}{\sqrt{2\omega_{k_i}}}\right):_b.
\eeq
Using the identity 
\beq
B^{C}_n(k_1\cdots k_n)=\sum_{J\subset[1,n]}\left(\prod_{j\in J}\frac{ b^{\dag}_{k_j}}{\sqrt{2\omega_{k_j}}}\right)\left(\prod_{j\in[1,n]\setminus J} \frac{b_{-k_j}}{\sqrt{2\omega_{k_j}}}\right)
\eeq
one finds the commutator
\bea
&&\hspace{-2cm}\left[\frac{b^\dag_{k\p}-b_{-k\p}}{\sqrt{2\omega_{k\p}}},B^{C}_n(k_1\cdots k_n)\right]\\
&&=-\frac{1}{2\omega_{k\p}}\sum_{J\subset[1,n]}\left[\sum_{j\p\in J}2\pi\delta(k_{j\p}+k\p)\prod_{j\in J\setminus j\p} \frac{b^{\dag}_{k_j}}{\sqrt{2\omega_{k_j}}}\prod_{j\in[1,n]\setminus J} \frac{b_{-k_j}}{\sqrt{2\omega_{k_j}}}\right.\nonumber\\
&&\hspace{2.5cm}\left.+\sum_{j\p\in [1,n]\setminus J}2\pi\delta(k_{j\p}+k\p)\prod_{j\in J} \frac{b^{\dag}_{k_j}}{\sqrt{2\omega_{k_j}}}\prod_{j\in[1,n]\setminus J\setminus j\p} \frac{b_{-k_j}}{\sqrt{2\omega_{k_j}}}\right]\nonumber\\
&&=-\frac{2}{2\omega_{k\p}}\sum_{j\p\in [1,n]}2\pi\delta(k_{j\p}+k\p)B^{C}_{n-1}(k_1\cdots\hat{k}_{j\p}\cdots k_n)\nonumber
\eea
and similarly the anticommutator
\bea
\left\{\frac{b^\dag_{k\p}+b_{-k\p}}{\sqrt{2\omega_{k\p}}},B^{C}_n(k_1\cdots k_n)\right\}&=&
2B^{C}_{n+1}(k_1\cdots k_n, k\p)\\
&+&\frac{2}{2\omega_{k\p}}\sum_{j\p\in [1,n]}2\pi\delta(k_{j\p}+k\p)B^{C}_{n-1}(k_1\cdots\hat{k}_{j\p}\cdots k_n).\nonumber
\eea
We will need the integrals of these identities, where the integral over $k\p$ is performed using the Dirac delta function
\bea
&&\left[\frac{1}{2}\pin{k\p}\tilde{g}_{k\p}(p_{n+1})\frac{\omega_{k\p}}{\omega_{p_{n+1}}}\frac{\left(b^\dag_{k\p}-b_{-k\p}\right)}{\sqrt{2\omega_{k\p}}},\pink{n-2m}\alpha_{nm}^{k_1\cdots k_{n-2m}} B^{C}_{n-2m}(k_1\cdots k_{n-2m})\right]\nonumber\\
&&=-\frac{1}{2\omega_{p_{n+1}}}\sum_{j\p=1}^{n-2m}\pink{n-2m}\tilde{g}_{-k_{j\p}}(p_{n+1})\alpha_{nm}^{k_1\cdots k_{n-2m}} B^{C}_{n-2m-1}(k_1\cdots\hat{k}_{j\p}\cdots k_{n-2m})
\eea
and
\bea
&&\left\{\frac{1}{2}\pin{k\p}\tilde{g}_{k\p}(p_{n+1})\frac{\left(b^\dag_{k\p}+b_{-k\p}\right)}{\sqrt{2\omega_{k_{n+1}}}},\pink{n-2m}\alpha_{nm}^{k_1\cdots k_{n-2m}} B^{C}_{n-2m}(k_1\cdots k_{n-2m})\right\}\nonumber\\
&&=\pink{n-2m}\pin{k\p}\tilde{g}_{k\p}(p_{n+1})\alpha_{nm}^{k_1\cdots k_{n-2m}} B^{C}_{n-2m+1}(k_1\cdots k_{n-2m},k\p)
\nonumber\\
&&+\sum_{j\p=1}^{n-2m}\pink{n-2m}\tilde{g}_{-k_{j\p}}(p_{n+1})\alpha_{nm}^{k_1\cdots k_{n-2m}} \frac{1}{2\omega_{k_{j\p}}}B^{C}_{n-2m-1}(k_1\cdots\hat{k}_{j\p}\cdots k_{n-2m})
\eea
for arbitrary matrices $\alpha_{nm}$.

We will define the matrices $\alpha_{nm}$ by
\beq
N^{C}_n(p_1\cdots p_n)=\left(\prod_{i=1}^n\sqrt{2\omega_{p_i}}\right)\sum_{m=0}^{\lfloor{\frac{n}{2}}\rfloor}\pink{n-2m}\alpha_{nm}^{k_1\cdots k_{n-2m}}B^{C}_{n-2m}(k_1\cdots k_{n-2m}).
\eeq
Then (\ref{guess}) implies
\bea
&&N^{C}_{n+1}(p_1\cdots p_{n+1})=\frac{1}{2}\left(\prod_{i=1}^{n+1}\sqrt{2\omega_{p_i}}\right)\sum_{m=0}^{\lfloor{\frac{n}{2}}\rfloor}\\
&&\left(\left\{\pin{k\p}\tilde{g}_{k\p}(p_{n+1})\frac{\left(b^\dag_{k\p}+b_{-k\p}\right)}{\sqrt{2\omega_{k_{n+1}}}},\pink{n-2m}\alpha_{nm}^{k_1\cdots k_{n-2m}} B^{C}_{n-2m}(k_1\cdots k_{n-2m})\right\}\right.\nonumber\\
&&\left.+\left[\pin{k\p}\tilde{g}_{k\p}(p_{n+1})\frac{\omega_{k\p}}{\omega_{p_{n+1}}}\frac{\left(b^\dag_{k\p}-b_{-k\p}\right)}{\sqrt{2\omega_{k\p}}},\pink{n-2m}\alpha_{nm}^{k_1\cdots k_{n-2m}} B^{C}_{n-2m}(k_1\cdots k_{n-2m})\right]\right)\nonumber\\
&=&\left(\prod_{i=1}^{n+1}\sqrt{2\omega_{p_i}}\right)\sum_{m=0}^{\lfloor{\frac{n}{2}}\rfloor}\left[\pink{n-2m}\pin{k\p}\tilde{g}_{k\p}(p_{n+1})\alpha_{nm}^{k_1\cdots k_{n-2m}} B^{C}_{n-2m+1}(k_1\cdots k_{n-2m},k\p)\right.\nonumber\\
&&\left.+\sum_{j\p=1}^{n-2m}\pink{n-2m}\left(\frac{1}{2\omega_{k_{j\p}}}-\frac{1}{2\omega_{p_{n+1}}}\right)\tilde{g}_{-k_{j\p}}(p_{n+1})\alpha_{nm}^{k_1\cdots k_{n-2m}}B^{C}_{n-2m-1}(k_1\cdots\hat{k}_{j\p}\cdots k_{n-2m})
\right].\nonumber
\eea
Summarizing, we find
\bea
&&\sum_{m=0}^{\lfloor{\frac{n+1}{2}}\rfloor}\pink{n-2m+1}\alpha_{n+1,m}^{k_1\cdots k_{n-2m+1}}B^{C}_{n-2m+1}(k_1\cdots k_{n-2m+1})\\
&&=\sum_{m=0}^{\lfloor{\frac{n}{2}}\rfloor}\left[\pink{n-2m}\pin{k\p}\tilde{g}_{k\p}(p_{n+1})\alpha_{nm}^{k_1\cdots k_{n-2m}} B^{C}_{n-2m+1}(k_1\cdots k_{n-2m},k\p)\right.\nonumber\\
&&\left.+\sum_{j\p=1}^{n-2m}\pink{n-2m}\left(\frac{1}{2\omega_{k_{j\p}}}-\frac{1}{2\omega_{p_{n+1}}}\right)\tilde{g}_{-k_{j\p}}(p_{n+1})\alpha_{nm}^{k_1\cdots k_{n-2m}}B^{C}_{n-2m-1}(k_1\cdots\hat{k}_{j\p}\cdots k_{n-2m})
\right]\nonumber\\
&&=\sum_{m=0}^{\lfloor{\frac{n}{2}}\rfloor}\pink{n-2m+1}\tilde{g}_{k_{n-2m+1}}(p_{n+1})\alpha_{nm}^{k_1\cdots k_{n-2m}} B^{C}_{n-2m+1}(k_1\cdots k_{n-2m+1})\nonumber\\
&&+\sum_{m=1}^{\lfloor{\frac{n}{2}}\rfloor+1}\sum_{j\p=1}^{n-2m+2}\pink{n-2m+2}\left(\frac{1}{2\omega_{k_{j\p}}}-\frac{1}{2\omega_{p_{n+1}}}\right)\nonumber\\
&&\times\tilde{g}_{-k_{j\p}}(p_{n+1})\alpha_{n,m-1}^{k_1\cdots k_{n-2m+2}}B^{C}_{n-2m+1}(k_1\cdots\hat{k}_{j\p}\cdots k_{n-2m+2})
\nonumber
\eea
where $\hat{k}_{j\p}$ indicates that $k_{j\p}$ is omitted.  Matching yields the recursion relation
\bea
\alpha_{n+1,m}^{k_1\cdots k_{n-2m+1}}&=&\tilde{g}_{k_{n-2m+1}}(p_{n+1})\alpha_{nm}^{k_1\cdots k_{n-2m}}\\
&&+\pin{k\p}\tilde{g}_{-k\p}(p_{n+1})\left(\frac{1}{2\omega_{k\p}}-\frac{1}{2\omega_{p_{n+1}}}\right)\sum_{j\p=1}^{n-2m+2}\alpha_{n,m-1}^{k_1\cdots k_{j\p-1}k\p k_{j\p}\cdots k_{n-2m+1}}.\nonumber
\eea

Symmetrizing we may write
\bea
N^{C}_n(p_1\cdots p_n)&=&\left(\prod_{i=1}^n\sqrt{2\omega_{p_i}}\right)\sum_{m=0}^{\lfloor{\frac{n}{2}}\rfloor}\pink{n-2m}\left(\prod_{i=1}^{n-2m}\tilde{g}_{k_i}(p_i)\right)a_{nm}B^{C}_{n-2m}(k_1\cdots k_{n-2m})\nonumber\\
&&\hspace{-1.5cm}\times\int\frac{d^m k\p}{(2\pi)^m}\prod_{i=1}^{m}\left(\tilde{g}_{-k\p_i}(p_{n-2m+2i-1})\tilde{g}_{k\p_i}(p_{n-2m+2i})\left(\frac{1}{2\omega_{k\p_i}}-\frac{1}{2\omega_{p_{n-2m+2i}}}\right)\right)
\eea
where again $a_{nm}$ satisfies (\ref{arec}) and so is given by (\ref{aeq}).   We therefore conclude
\bea
:\phi^j_C(x):_a&=&\sum_{m=0}^{\lfloor{\frac{j}{2}}\rfloor}I_C^m(x)\pink{j-2m}\left(\prod_{i=1}^{j-2m}g_{k_i}(x)\right)\frac{j!}{m!(j-2m)!}B^{C}_{j-2m}(k_1\cdots k_{j-2m})\nonumber\\
&=&\sum_{m=0}^{\lfloor{\frac{j}{2}}\rfloor}\frac{j!}{m!(j-2m)!}I_C^m(x):\phi^{j-2m}_C(x):_b\nonumber\\
I_C(x)&=&\frac{1}{2}\pin{k}g_{-k}(x)\hat{g}_k(x)\hsp
\hat{g}_k(x)=\pin{p}e^{-ipx}\tilde{g}_k(p)\left(\frac{1}{2\omega_{k}}-\frac{1}{2\omega_p}\right).
\eea
While the algebra leading up to our result seemed more complicated than in the case of the bound states, our final result is essentially the same.  The only difference is that $I_C$ is integrated over normal modes $k$.  However, even in the case of breathers, there will be a sum over breather modes $i$, and so this distinction is superficial.

\section{Remarks}

We have found that plane wave normal ordering can be converted into normal mode normal ordering by following a simple rule, playing the role of Wick's theorem.  After decomposing a product of $n$ fields into products of $j$ field components, where each component corresponds to a set of normal modes, the components can be decomposed by summing over all possible contractions.  For each contraction one replaces the pair of field components with the difference between the inverse plane wave energy $\omega_p$ and inverse normal mode energy, suitably normalized over the spectrum.  Intuitively the first term arises from eliminating the plane wave normal ordering and the second from imposing the normal mode normal ordering.  Of course with no normal ordering at all, one expects divergences.  However the difference between these two energies is, when suitably averaged, quite small and thus all expressions are finite given either normal ordering scheme.  Once we go beyond scalar theories and 1+1 dimensions there will be other divergences which must be regularized and renormalized.

In Ref.~\cite{meduestato} the conversion between normal orderings was the most complicated part of the perturbation theory treatment of the one soliton sector.  Now that we have treated this problem at all orders, and in a much more general class of theories, we expect that it will be easier to extend that calculation to two loops or beyond.  The results could then be compared with Refs.~\cite{dhn2loops,luther,vega}.  However it is still not obvious that the solution to the zero mode problem in Ref.~\cite{meduestato} also solves the problem at higher loops.  If it does not, then it may be necessary to use other formalisms such as that of \cite{christleecc} and \cite{sakitacc}.

To go beyond perturbation theory, we will eventually need supersymmetry.  In this context, coherent states have been constructed in Refs.~\cite{firrotta1,firrotta2}.  This will require a fermionic generalization of the Wick's theorem found here.  Perhaps the generalized Wick's theorem of Ref.~\cite{diosi} can provide an efficient derivation.

The recent discovery of spectral walls \cite{wall} caused by transitions between breather and continuum states has rekindled interest in kink scattering \cite{lankink,campos}.  The treatment of this phenomenon has so far been largely classical.  While the current methodology is most straightforwardly applied to the one kink sector, it could nonetheless allow an understanding of the role played by breathers in fully quantum scattering.  In particular the scattering of a kink with a plane wave or wave packet could be treated in the one kink sector.  For this an interaction picture generalization of the results above may be desirable.

\section* {Acknowledgement}

\noindent
We thank Hengyuan Guo for a careful reading of this manuscript.  JE is supported by the CAS Key Research Program of Frontier Sciences grant QYZDY-SSW-SLH006 and the NSFC MianShang grants 11875296 and 11675223.   JE also thanks the Recruitment Program of High-end Foreign Experts for support.


\begin{thebibliography}{99}

\bibitem{delfino}
  G.~Delfino, W.~Selke and A.~Squarcini,
  ``Vortex mass in the three-dimensional $O(2)$ scalar theory,''
  Phys.\ Rev.\ Lett.\  {\bf 122} (2019) no.5,  050602
  doi:10.1103/PhysRevLett.122.050602
  [arXiv:1808.09276 [cond-mat.stat-mech]].

\bibitem{davies}
  D.~Davies,
  ``Quantum Solitons in any Dimension: Derrick's Theorem v. AQFT,''
  arXiv:1907.10616 [hep-th].


\bibitem{hepp}
  K.~Hepp,
  ``The Classical Limit for Quantum Mechanical Correlation Functions,''
  Commun.\ Math.\ Phys.\  {\bf 35} (1974) 265.
  doi:10.1007/BF01646348

\bibitem{taylor78}
  J.~G.~Taylor,
  ``Solitons as Infinite Constituent Bound States,''
  Annals Phys.\  {\bf 115} (1978) 153.
  doi:10.1016/0003-4916(78)90179-3


\bibitem{sato}
J.~Sato and T.~Yumibayashi,
``Quantum-classical correspondence via coherent state in integrable field theory,''
[arXiv:1811.03186 [quant-ph]].


\bibitem{cahill76}
K.~E.~Cahill, A.~Comtet and R.~Glauber,
``Mass Formulas for Static Solitons,''
Phys. Lett. B \textbf{64} (1976), 283-285
doi:10.1016/0370-2693(76)90202-1


\bibitem{christleecc}
N.~Christ and T.~Lee,
``Quantum Expansion of Soliton Solutions,''
Phys. Rev. D \textbf{12} (1975), 1606
doi:10.1103/PhysRevD.12.1606

\bibitem{sakitacc}
J.~L.~Gervais and B.~Sakita,
``Extended Particles in Quantum Field Theories,''
Phys. Rev. D \textbf{11} (1975), 2943
doi:10.1103/PhysRevD.11.2943

\bibitem{meduestato}
J.~Evslin,
``Constructing Quantum Soliton States Despite Zero Modes,''
[arXiv:2006.02354 [hep-th]].

\bibitem{dhn2}
  R.~F.~Dashen, B.~Hasslacher and A.~Neveu,
  ``Nonperturbative Methods and Extended Hadron Models in Field Theory 2. Two-Dimensional Models and Extended Hadrons,''
  Phys.\ Rev.\ D {\bf 10} (1974) 4130.
 doi:10.1103/PhysRevD.10.4130

\bibitem{rajaraman}
  R.~Rajaraman,
  ``Some Nonperturbative Semiclassical Methods in Quantum Field Theory: A Pedagogical Review,''
  Phys.\ Rept.\  {\bf 21} (1975) 227.
  doi:10.1016/0370-1573(75)90016-2

\bibitem{aguirre}
A.~Aguirre and G.~Flores-Hidalgo,
``A note on one-loop soliton quantum mass corrections,''
Mod. Phys. Lett. A \textbf{33} (2020), 2050102
doi:10.1142/S0217732320501023
[arXiv:1912.13051 [hep-th]].

\bibitem{mekink}
J.~Evslin,
``Manifestly Finite Derivation of the Quantum Kink Mass,''
JHEP \textbf{11} (2019), 161
doi:10.1007/JHEP11(2019)161
[arXiv:1908.06710 [hep-th]].

\bibitem{memassa}
J.~Evslin,
``Well-defined quantum soliton masses without supersymmetry,''
Phys. Rev. D \textbf{101} (2020) no.6, 065005
doi:10.1103/PhysRevD.101.065005
[arXiv:2002.12523 [hep-th]].

\bibitem{rebhan}
  A.~Rebhan and P.~van Nieuwenhuizen,
  ``No saturation of the quantum Bogomolnyi bound by two-dimensional supersymmetric solitons,''
  Nucl.\ Phys.\ B {\bf 508} (1997) 449
  doi:10.1016/S0550-3213(97)00625-1, 10.1016/S0550-3213(97)80021-1
 [hep-th/9707163].




\bibitem{sg}
H.~Guo and J.~Evslin,
``Finite derivation of the one-loop sine-Gordon soliton mass,''
JHEP \textbf{02} (2020), 140
doi:10.1007/JHEP02(2020)140
[arXiv:1912.08507 [hep-th]].

\bibitem{diosi}
L.~Diósi,
``Wick theorem for all orderings of canonical operators,''
J. Phys. A \textbf{51} (2018) no.36, 365201
doi:10.1088/1751-8121/aad0a6
[arXiv:1712.08811 [quant-ph]].


\bibitem{dhn2loops}
R.~F.~Dashen, B.~Hasslacher and A.~Neveu,
``The Particle Spectrum in Model Field Theories from Semiclassical Functional Integral Techniques,''
Phys. Rev. D \textbf{11} (1975), 3424
doi:10.1103/PhysRevD.11.3424

\bibitem{luther}
A.~Luther,
``Eigenvalue spectrum of interacting massive fermions in one-dimension,''
Phys. Rev. B \textbf{14} (1976), 2153-2159
doi:10.1103/PhysRevB.14.2153

\bibitem{vega}
H.~de Vega,
``Two-Loop Quantum Corrections to the Soliton Mass in Two-Dimensional Scalar Field Theories,''
Nucl. Phys. B \textbf{115} (1976), 411-428
doi:10.1016/0550-3213(76)90497-1

\bibitem{firrotta1}
M.~Bianchi and M.~Firrotta,
``DDF operators, open string coherent states and their scattering amplitudes,''
Nucl. Phys. B \textbf{952} (2020), 114943
doi:10.1016/j.nuclphysb.2020.114943
[arXiv:1902.07016 [hep-th]].

\bibitem{firrotta2}
A.~Aldi and M.~Firrotta,
``String coherent vertex operators of Neveu-Schwarz and Ramond states,''
Nucl. Phys. B \textbf{955} (2020), 115050
doi:10.1016/j.nuclphysb.2020.115050
[arXiv:1912.06177 [hep-th]].

\bibitem{wall}
C.~Adam, K.~Oles, T.~Romanczukiewicz and A.~Wereszczynski,
``Spectral Walls in Soliton Collisions,''
Phys. Rev. Lett. \textbf{122} (2019) no.24, 241601
doi:10.1103/PhysRevLett.122.241601
[arXiv:1903.12100 [hep-th]].

\bibitem{lankink}
Y.~Zhong, X.~L.~Du, Z.~C.~Jiang, Y.~X.~Liu and Y.~Q.~Wang,
``Collision of two kinks with inner structure,''
JHEP \textbf{02} (2020), 153
doi:10.1007/JHEP02(2020)153
[arXiv:1906.02920 [hep-th]].

\bibitem{campos}
J.~G.~Campos and A.~Mohammadi,
``Interaction between kinks and antikinks with double long-range tails,''
[arXiv:2006.01956 [hep-th]].

\end{thebibliography}

\begin{thebibliography}{99}

\bibitem{hepp}
  K.~Hepp,
  ``The Classical Limit for Quantum Mechanical Correlation Functions,''
  Commun.\ Math.\ Phys.\  {\bf 35} (1974) 265.
  doi:10.1007/BF01646348

\bibitem{sato}
J.~Sato and T.~Yumibayashi,
``Quantum-classical correspondence via coherent state in integrable field theory,''
[arXiv:1811.03186 [quant-ph]].

\bibitem{taylor78}
  J.~G.~Taylor,
  ``Solitons as Infinite Constituent Bound States,''
  Annals Phys.\  {\bf 115} (1978) 153.
  doi:10.1016/0003-4916(78)90179-3

\bibitem{delfino}
  G.~Delfino, W.~Selke and A.~Squarcini,
  ``Vortex mass in the three-dimensional $O(2)$ scalar theory,''
  Phys.\ Rev.\ Lett.\  {\bf 122} (2019) no.5,  050602
  doi:10.1103/PhysRevLett.122.050602
  [arXiv:1808.09276 [cond-mat.stat-mech]].

\bibitem{davies}
  D.~Davies,
  ``Quantum Solitons in any Dimension: Derrick's Theorem v. AQFT,''
  arXiv:1907.10616 [hep-th].


\bibitem{colemansg}
  S.~R.~Coleman,
  ``The Quantum Sine-Gordon Equation as the Massive Thirring Model,''
  Phys.\ Rev.\ D {\bf 11} (1975) 2088.
  doi:10.1103/PhysRevD.11.2088


\bibitem{mandelstamsol}
  S.~Mandelstam,
  ``Soliton Operators for the Quantized Sine-Gordon Equation,''
  Phys.\ Rev.\ D {\bf 11} (1975) 3026.
  doi:10.1103/PhysRevD.11.3026

\bibitem{sw2}
  N.~Seiberg and E.~Witten,
  ``Electric - magnetic duality, monopole condensation, and confinement in N=2 supersymmetric Yang-Mills theory,''
  Nucl.\ Phys.\ B {\bf 426} (1994) 19
   Erratum: [Nucl.\ Phys.\ B {\bf 430} (1994) 485]
  doi:10.1016/0550-3213(94)90124-4, 10.1016/0550-3213(94)00449-8
  [hep-th/9407087].

\bibitem{thooftconf}
  G.~'t Hooft,
  ``Topology of the Gauge Condition and New Confinement Phases in Nonabelian Gauge Theories,''
  Nucl.\ Phys.\ B {\bf 190} (1981) 455.
  doi:10.1016/0550-3213(81)90442-9

\bibitem{mandelconf}
  S.~Mandelstam,
  ``Vortices and Quark Confinement in Nonabelian Gauge Theories,''
  Phys.\ Rept.\  {\bf 23} (1976) 245.
  doi:10.1016/0370-1573(76)90043-0

\bibitem{rebsol}
A.~Aguirre and G.~Flores-Hidalgo,
``A note on one-loop soliton quantum mass corrections,''
Mod. Phys. Lett. A \textbf{33} (2020), 2050102
doi:10.1142/S0217732320501023
[arXiv:1912.13051 [hep-th]].

\bibitem{dhn2}
  R.~F.~Dashen, B.~Hasslacher and A.~Neveu,
  ``Nonperturbative Methods and Extended Hadron Models in Field Theory 2. Two-Dimensional Models and Extended Hadrons,''
  Phys.\ Rev.\ D {\bf 10} (1974) 4130.
 doi:10.1103/PhysRevD.10.4130

\bibitem{rajaraman}
  R.~Rajaraman,
  ``Some Nonperturbative Semiclassical Methods in Quantum Field Theory: A Pedagogical Review,''
  Phys.\ Rept.\  {\bf 21} (1975) 227.
  doi:10.1016/0370-1573(75)90016-2

\bibitem{physrept04}
  A.~S.~Goldhaber, A.~Rebhan, P.~van Nieuwenhuizen and R.~Wimmer,
  ``Quantum corrections to mass and central charge of supersymmetric solitons,''
  Phys.\ Rept.\  {\bf 398} (2004) 179
  doi:10.1016/j.physrep.2004.05.001
  [hep-th/0401152].

\bibitem{rebhan}
  A.~Rebhan and P.~van Nieuwenhuizen,
  ``No saturation of the quantum Bogomolnyi bound by two-dimensional supersymmetric solitons,''
  Nucl.\ Phys.\ B {\bf 508} (1997) 449
  doi:10.1016/S0550-3213(97)00625-1, 10.1016/S0550-3213(97)80021-1
 [hep-th/9707163].

\bibitem{dhn1}
R.~F.~Dashen, B.~Hasslacher and A.~Neveu,
``Nonperturbative Methods and Extended Hadron Models in Field Theory 1. Semiclassical Functional Methods,''
Phys. Rev. D \textbf{10} (1974), 4114
doi:10.1103/PhysRevD.10.4114

\bibitem{mestato}
J.~Evslin,
``The Ground State of the Sine-Gordon Soliton,''
[arXiv:2003.11384 [hep-th]].

\bibitem{friedrichscont}
K.O. Friedrichs,
``On the perturbation of continuous spectra,''
Commun.Pure Appl.Math. 1 (1948) 361-406
doi:10.1002/cpa.3160010404

\bibitem{callangross}
C.~G.~Callan, Jr. and D.~J.~Gross,
``Quantum Perturbation Theory of Solitons,''
Nucl. Phys. B \textbf{93} (1975), 29-55
doi:10.1016/0550-3213(75)90150-9

\bibitem{coleman2d}
S.~R.~Coleman,
``There are no Goldstone bosons in two-dimensions,''
Commun. Math. Phys. \textbf{31} (1973), 259-264
doi:10.1007/BF01646487

\bibitem{mekink}
J.~Evslin,
``Manifestly Finite Derivation of the Quantum Kink Mass,''
JHEP \textbf{11} (2019), 161
doi:10.1007/JHEP11(2019)161
[arXiv:1908.06710 [hep-th]].

\bibitem{memassa}
J.~Evslin and B.~Zhang,
``Well-defined quantum soliton masses without supersymmetry,''
Phys. Rev. D \textbf{101} (2020) no.6, 065005
doi:10.1103/PhysRevD.101.065005
[arXiv:2002.12523 [hep-th]].



\bibitem{sg}
H.~Guo and J.~Evslin,
``Finite derivation of the one-loop sine-Gordon soliton mass,''
JHEP \textbf{02} (2020), 140
doi:10.1007/JHEP02(2020)140
[arXiv:1912.08507 [hep-th]].


\bibitem{flugge}
S. Fl\"ugge,
``Practical Quantum Mechanics,"
Springer-Verlag Berlin Heidelberg (1999),
doi:10.1007/978-3-642-61995-3

\bibitem{vanhove1}
L.~Van Hove,
``Energy corrections and persistent perturbation effects in continuous spectra,''
Physica \textbf{21} (1955), 901-923
doi:10.1016/S0031-8914(55)92832-9  
  
\bibitem{vanhove2}
L.~Van Hove,
``Energy corrections and persistent perturbation effects in continuous spectra. II: The perturbed stationary states,''
Physica \textbf{22} (1956), 343-354
doi:10.1016/S0031-8914(56)80046-3

\bibitem{sakitacc}
J.~L.~Gervais and B.~Sakita,
``Extended Particles in Quantum Field Theories,''
Phys. Rev. D \textbf{11} (1975), 2943
doi:10.1103/PhysRevD.11.2943


\bibitem{stuart}
D.~Stuart, M.A.,
``Hamiltonian quantization of solitons in the $\phi^4_{1+1}$ quantum field theory. I. The semiclassical mass shift,''
[arXiv:1904.02588 [math-ph]].

\bibitem{firrotta1}
M.~Bianchi and M.~Firrotta,
``DDF operators, open string coherent states and their scattering amplitudes,''
Nucl. Phys. B \textbf{952} (2020), 114943
doi:10.1016/j.nuclphysb.2020.114943
[arXiv:1902.07016 [hep-th]].

\bibitem{firrotta2}
A.~Aldi and M.~Firrotta,
``String coherent vertex operators of Neveu-Schwarz and Ramond states,''
Nucl. Phys. B \textbf{955} (2020), 115050
doi:10.1016/j.nuclphysb.2020.115050
[arXiv:1912.06177 [hep-th]].


\bibitem{dhn2loops}
R.~F.~Dashen, B.~Hasslacher and A.~Neveu,
``The Particle Spectrum in Model Field Theories from Semiclassical Functional Integral Techniques,''
Phys. Rev. D \textbf{11} (1975), 3424
doi:10.1103/PhysRevD.11.3424

\bibitem{luther}
A.~Luther,
``Eigenvalue spectrum of interacting massive fermions in one-dimension,''
Phys. Rev. B \textbf{14} (1976), 2153-2159
doi:10.1103/PhysRevB.14.2153

\bibitem{vega}
H.~de Vega,
``Two-Loop Quantum Corrections to the Soliton Mass in Two-Dimensional Scalar Field Theories,''
Nucl. Phys. B \textbf{115} (1976), 411-428
doi:10.1016/0550-3213(76)90497-1

\bibitem{mussardo}
G.~Mussardo, V.~Riva and G.~Sotkov,
``Semiclassical scaling functions of sine-Gordon model,''
Nucl. Phys. B \textbf{699} (2004), 545-574
doi:10.1016/j.nuclphysb.2004.08.004
[arXiv:hep-th/0405139 [hep-th]].


\end{thebibliography}

\begin{thebibliography}{99}

\bibitem{dhn2}
  R.~F.~Dashen, B.~Hasslacher and A.~Neveu,
  ``Nonperturbative Methods and Extended Hadron Models in Field Theory 2. Two-Dimensional Models and Extended Hadrons,''
  Phys.\ Rev.\ D {\bf 10} (1974) 4130.
  doi:10.1103/PhysRevD.10.4130
  
\bibitem{dhnsg}
  R.~F.~Dashen, B.~Hasslacher and A.~Neveu,
  ``The Particle Spectrum in Model Field Theories from Semiclassical Functional Integral Techniques,''
  Phys.\ Rev.\ D {\bf 11} (1975) 3424.
  doi:10.1103/PhysRevD.11.3424


\bibitem{colemansg}
  S.~R.~Coleman,
  ``The Quantum Sine-Gordon Equation as the Massive Thirring Model,''
  Phys.\ Rev.\ D {\bf 11} (1975) 2088.
  doi:10.1103/PhysRevD.11.2088

\bibitem{luther}
  A.~Luther,
  ``Eigenvalue spectrum of interacting massive fermions in one-dimension,''
  Phys.\ Rev.\ B {\bf 14} (1976) 2153.
  doi:10.1103/PhysRevB.14.2153


\bibitem{re}
  A.Rebhan and P.Van Nieuwenhuizen,
  ``No saturation of the quantum Bogomolnyi bound by two-dimensional supersymmetric solitons,''
  [hep-th/9707163]

\bibitem{mekink}
 J.~Evslin,
  ``Manifestly Finite Derivation of the Quantum Kink Mass,''
  JHEP {\bf 1911} (2019) 161
  doi:10.1007/JHEP11(2019)161
  [arXiv:1908.06710 [hep-th]].

\bibitem{hui}
  H.~Liu, Y.~Zhou and J.~Evslin,
  ``Ground States of the $\phi^4$ Double-Well QFT,''
  arXiv:1909.04946 [hep-th].

\bibitem{taylor78}
  J.~G.~Taylor,
  ``Solitons as Infinite Constituent Bound States,''
  Annals Phys.\  {\bf 115} (1978) 153.
  doi:10.1016/0003-4916(78)90179-3


\bibitem{hepp}
  K.~Hepp,
  ``The Classical Limit for Quantum Mechanical Correlation Functions,''
  Commun.\ Math.\ Phys.\  {\bf 35} (1974) 265.
  doi:10.1007/BF01646348






\bibitem{flugge}
S. Fl\"ugge,
``Practical Quantum Mechanics,"
Springer-Verlag Berlin Heidelberg (1999),
doi:10.1007/978-3-642-61995-3

\bibitem{johnson73}
  J.~D.~Johnson, S.~Krinsky and B.~M.~McCoy,
  ``Vertical-Arrow Correlation Length in the Eight-Vertex Model and the Low-Lying Excitations of the X-Y-Z Hamiltonian,''
  Phys.\ Rev.\ A {\bf 8} (1973) 2526.

\bibitem{ft}
  L.~D.~Faddeev, L.~A.~Takhtajan and V.~E.~Zakharov,
  ``Complete description of solutions of the Sine-Gordon equation,''
  Dokl.\ Akad.\ Nauk Ser.\ Fiz.\  {\bf 219} (1974) 1334
   [Sov.\ Phys.\ Dokl.\  {\bf 19} (1975) 824].


\bibitem{mandelop}
  S.~Mandelstam,
  ``Soliton Operators for the Quantized Sine-Gordon Equation,''
  Phys.\ Rev.\ D {\bf 11} (1975) 3026.
  doi:10.1103/PhysRevD.11.3026

\bibitem{sw2}
  N.~Seiberg and E.~Witten,
  ``Electric - magnetic duality, monopole condensation, and confinement in N=2 supersymmetric Yang-Mills theory,''
  Nucl.\ Phys.\ B {\bf 426} (1994) 19
   Erratum: [Nucl.\ Phys.\ B {\bf 430} (1994) 485]
  doi:10.1016/0550-3213(94)90124-4, 10.1016/0550-3213(94)00449-8
  [hep-th/9407087].


\end{thebibliography}
\end{document}